\documentclass[ALICE,manyauthors]{cernphprep}
\usepackage{hyperref}
\usepackage{bm}
\def\pt{$p_{\rm T}$}
\def\mpt{$\left<p_{\rm T}\right>$}

\def\mdndeta{$\left<{\rm d}N_{\rm ch}/{\rm d}\eta\right>$}
\def\roots{$\sqrt{s}$}
\def\rootsnn{$\sqrt{s_{\rm NN}}$}
\def\rspt{$\sqrt{C_m}/M(p_{\rm T})_m$}
\def\rsptinc{$\sqrt{C}/M(p_{\rm T})$}
\def\mnpart{$\langle N_{\rm part} \rangle$}
\begin{document}
%
%
%
\begin{titlepage}
\PHnumber{2014-169}                 
\PHdate{15 July 2014}              
%
%
\title{Event-by-event mean $\bm{p}_{\mathbf{T}}$~fluctuations \\ in pp and Pb--Pb collisions at the LHC}
\ShortTitle{Event-by-event mean \pt~fluctuations at the LHC}   
%
\Collaboration{ALICE Collaboration%
         \thanks{See Appendix~\ref{app:collab} for the list of collaboration
                      members}
}
\ShortAuthor{ALICE Collaboration}      
\begin{abstract}
Event-by-event fluctuations of the mean transverse momentum of charged particles
produced in pp collisions at \roots~$=$~0.9, 2.76 and 7~TeV, and Pb--Pb collisions
at \rootsnn~$=$~2.76~TeV are studied as a function of the charged-particle multiplicity 
using the ALICE detector at the LHC.
Dynamical fluctuations indicative of correlated particle emission are observed in all systems.
The results in pp collisions show little dependence on collision energy.
The Monte Carlo event generators PYTHIA and PHOJET are in qualitative agreement 
with the data.
Peripheral Pb--Pb data exhibit a similar multiplicity dependence as that observed in pp.
In central Pb--Pb, the results deviate from this trend, featuring a significant reduction 
of the fluctuation strength. 
The results in Pb--Pb are in qualitative agreement with previous measurements in Au--Au 
at lower collision energies and with expectations from models that incorporate collective 
phenomena.
\end{abstract}
\end{titlepage}
\setcounter{page}{2}
%
%

\section{Introduction}

The study of event-by-event fluctuations was proposed as a probe of the 
properties of the hot and dense matter generated in high-energy heavy-ion 
collisions~\cite{jeonkoch,mrow,stodolsky,shuryak,stephanov1,stephanov2,heiselberg,dumitru,fodor}. 
The occurrence of a phase transition from the Quark-Gluon Plasma to a Hadron Gas or the existence 
of a critical point in the phase diagram of strongly interacting matter 
may go along with critical fluctuations of thermodynamic quantities such as temperature. 
This could be reflected in dynamical event-by-event fluctuations of the mean transverse momentum 
$\left(  \left<p_{\rm T}\right> \right)$ of final-state charged particles.

Event-by-event \mpt~fluctuations have been studied in nucleus-nucleus (A--A) collisions 
at the Super Proton Synchrotron (SPS)~\cite{na49-1,ceres-1,na49-2,ceres-2,na49-3} 
and at the Relativistic Heavy-Ion Collider (RHIC)~\cite{phenix-1,phenix-2,star-1,star-2,star-3,star-4}, 
where dynamical fluctuations have been observed. 
Fluctuations of \mpt~were found to decrease with collision centrality, 
as generally expected in a dilution scenario caused by superposition 
of partially independent particle-emitting sources. 
In detail, however, deviations from a simple superposition
scenario have been reported. In particular, with respect to a reference representing 
independent superposition -- i.e.\,a decrease of fluctuations according to \mdndeta$^{-0.5}$,
where \mdndeta~is the average charged-particle density in a given interval
of collision centrality and pseudorapidity ($\eta$)~--
the observed fluctuations increase sharply from peripheral
to semi-peripheral collisions, followed by a shallow decrease towards central collisions~\cite{star-2}.
A number of possible mechanisms have been proposed to explain this behavior,
such as string percolation~\cite{perco} or the onset of thermalization and collectivity~\cite{volo-1,gavin}, 
but no strong connection to critical behavior could be made. 
It was recently suggested \cite{moschelli,moschelli-2} that initial state density 
fluctuations~\cite{alver} could affect the final state transverse momentum correlations and 
their centrality dependence.

Fluctuations of \mpt~arise from many kinds of correlations among the
\pt~of the final-state particles, such as resonance decays, jets, or quantum correlations.
To account for these contributions from conventional mechanisms similar studies can be 
performed in pp, where such correlations are also present. The results from pp could thus
be used to construct a model-independent baseline to search for non-trivial fluctuations in A--A 
which manifest themselves in a modification of the fluctuation pattern with respect
to the pp reference.

In this paper, we present results of a multiplicity-dependent study of event-by-event 
\mpt~fluctuations of charged particles in pp collisions 
at \roots~$=$~0.9, 2.76 and 7 TeV, and Pb--Pb collisions at \rootsnn~$=$~2.76~TeV  
measured with ALICE at the LHC. 
The experimental data are compared to different Monte Carlo (MC) event generators.

\section{ALICE detector and data analysis}

The data used in this analysis were collected with the ALICE detector at the CERN Large Hadron Collider 
(LHC) ~\cite{lhc} during the Pb--Pb run in 2010 and the pp runs in 2010 and 2011. 

For a detailed description of the ALICE apparatus see~\cite{alice}. 
The analysis is based on $19\times10^6$ Pb--Pb events at \rootsnn~$=$~2.76~TeV, 
and $6.9\times10^6$, $66\times10^6$ and $290\times10^6$ pp events at \roots~$=$~0.9, 2.76 and 7 TeV,
respectively. The standard ALICE coordinate system is used, in which the nominal interaction point 
is the origin of a right-handed Cartesian coordinate system. The $z$-axis is along the beam pipe,
the $x$-axis points towards the center of the LHC, $\varphi$ is the azimuthal angle around the 
$z$-axis and $\theta$ is the polar angle with respect to this axis. The detectors in the central barrel
of the experiment are operated inside a solenoidal magnetic field with $B=0.5$~T. 
About half of the Pb--Pb data set was recorded with negative ($B_z<0$) and positive ($B_z>0$) 
field polarity, respectively.

A minimum bias (MB) trigger condition was applied to select collision events.
In pp, this trigger was defined by at least one hit in the Silicon Pixel Detector (SPD) 
or in one of the two forward scintillator systems 
VZERO-A ($2.8 < \eta < 5.1$) and VZERO-C ($-3.7 < \eta < -1.7$). 
In Pb--Pb, the MB trigger condition is defined as a coincidence of 
hits in both VZERO detectors.

In this analysis, the Time Projection Chamber (TPC)~\cite{tpc} is used for charged-particle tracking 
in $|\eta|<0.8$. In the momentum range selected for this analysis, $0.15<p_{\rm T}<2$~GeV/$c$, 
the momentum resolution $\sigma(p_{\rm T})/p_{\rm T}$ is better than 2\%. 
The tracking efficiency is larger than 90\% for $p_{\rm T}>0.3$~GeV/$c$
and drops to about 70\% at $p_{\rm T}=0.15$~GeV/$c$.

Primary vertex information is obtained from both the Inner Tracking System (ITS) and the TPC. 
Events are used in the analysis when at least one accepted charged-particle track contributes to the 
primary vertex reconstruction. 
It is required that the reconstructed vertex is 
within $\pm10$~cm from the nominal interaction point along the beam direction 
to ensure a uniform pseudo-rapidity acceptance within the TPC. 
Additionaly, the event vertex is reconstructed using only TPC tracks. 
The event is rejected if the $z$-position of that vertex is different by more
than 10~cm from that of the standard procedure.

In Pb--Pb, at least 10 reconstructed tracks inside the acceptance are required. 
The contamination by non-hadronic interactions 
is negligible in the event sample that fulfills the aforementioned selection criteria. 
The centrality in Pb--Pb is estimated from the signal in the VZERO detectors using the procedure described 
in~\cite{dndeta,centr-new}.

The charged-particle tracks used for this analysis 
are required to have at least 70 out of a maximum of 159 reconstructed space points in the TPC, 
and the maximum $\chi^2$ per space point in the TPC from the momentum fit must be less than 4. 
Daughter tracks from reconstructed secondary weak-decay topologies ({\em kinks}) are rejected. 
The distance of closest approach (DCA) of the extrapolated trajectory to the primary vertex position 
is restricted to less than 3.2~cm along the beam direction and less than 2.4~cm in the transverse plane. 
The number of tracks in an event that are accepted by these selection criteria is denoted by $N_{\rm acc}$.

Event-by-event measurements of the mean transverse momentum are subject to the finite reconstruction 
efficiency of the detector. While efficiency corrections can be applied on a statistical basis 
to derive the inclusive \mpt~of charged particles in a given kinematic acceptance range, 
such an approach is not adequate for event-by-event studies. The event-by-event mean transverse momentum
is therefore approximated by the mean value $M_{\rm EbE}(p_{\rm T})_k$ of the transverse momenta 
$p_{{\rm T},i}$ of the $N_{\rm acc},_k$ accepted charged particles in event $k$:
\begin{equation}
  M_{\rm EbE}(p_{\rm T})_k=\frac{1}{N_{\rm acc},_k} \sum_{i=1}^{N_{\rm acc},_k} p_{{\rm T},i} \ .
\end{equation}

Event-by-event fluctuations of $M_{\rm EbE}(p_{\rm T})_k$ in heavy-ion collisions 
are composed of statistical and dynamical contributions. 
The two-particle transverse momentum correlator 
$C = \langle \Delta p_{{\rm T},i}, \Delta p_{{\rm T},j} \rangle$ is a measure of the 
dynamical component $\sigma_{\rm dyn}^2$ of these fluctuations and therefore well suited for an 
event-by-event analysis~\cite{ceres-2,star-2,voloshin}. 
The correlator $C_{m}$ is the mean of covariances of all pairs of particles $i$ and $j$ in the same event 
with respect to the inclusive $M(p_{\rm T})_m$ in a given multiplicity class $m$ and is defined as
\begin{equation}
  C_m = \frac{1}{\sum_{k=1}^{n_{{\rm ev},m}}{N_{k}^{\rm pairs}}} \cdot \sum_{k=1}^{n_{{\rm ev},m}} \sum_{i=1}^{N_{\rm acc},_{k}} \sum_{j=i+1}^{N_{\rm acc},_{k}} (p_{{\rm T},i} - M(p_{\rm T})_m ) \cdot (p_{{\rm T},j} - M(p_{\rm T})_m) \ ,
  \label{eq:correlator}
\end{equation}
where $n_{{\rm ev},m}$ is the number of events in multiplicity class $m$, 
$N_{k}^{\rm pairs} = 0.5 \cdot N_{\rm acc},_{k} \cdot (N_{\rm acc},_{k}-1)$ is the number of particle pairs 
in event $k$ and $M(p_{\rm T})_m$ is the average $p_{\rm T}$ of all tracks in all events of class $m$:
\begin{equation}
  M(p_{\rm T})_m=\frac{1}{\sum_{k=1}^{n_{{\rm ev},m}}N_{\rm acc},_k} \sum_{k=1}^{n_{{\rm ev},m}}\sum_{i=1}^{N_{\rm acc},_k} p_{{\rm T},i}=\frac{1}{\sum_{k=1}^{n_{{\rm ev},m}}N_{\rm acc},_k} \sum_{k=1}^{n_{{\rm ev},m}} N_{\rm acc},_k \cdot M_{\rm EbE}(p_{\rm T})_k \ .
\end{equation}
By construction, $C_m$ vanishes in the case of uncorrelated particle emission, when only statistical fluctuations are present.

The results in this paper are presented in terms of the dimensionless 
ratio \rspt~which quantifies the strength of the dynamical fluctuations in units of the average 
transverse momentum $M(p_{\rm T})_m$ in the multiplicity class $m$.

The correlator is computed in intervals of the event multiplicity $N_{\rm acc}$. In pp collisions, 
intervals of $\Delta N_{\rm acc}$~=~1 are used for the calculation of $C_m$ and $M(p_{\rm T})_m$. 
In Pb--Pb collisions, $C_m$ is calculated in the multiplicity intervals $\Delta N_{\rm acc}$~=~10 
for $N_{\rm acc} < 200$, $\Delta N_{\rm acc}$~=~25 for $200 \leq N_{\rm acc} < 1000$ and 
$\Delta N_{\rm acc}$~=~100 for $N_{\rm acc} \geq$~1000. 
To account for the steep increase of $M(p_{\rm T})_m$ with multiplicity in peripheral collisions, 
the calculation of the correlator in~(\ref{eq:correlator}) uses values for 
$M(p_{\rm T})_m$ which are calculated in bins of $\Delta N_{\rm acc}$~=~1 for $N_{\rm acc}<1000$. 
At higher multiplicities, $M(p_{\rm T})$ changes only moderately and $M(p_{\rm T})_m$ is 
calculated in the same intervals as $C_m$, i.e.\,$\Delta N_{\rm acc}$~=~100.

Additionally, the Pb--Pb data are also analyzed in 5\% intervals of the collision centrality. 
The results are shown in bins of the mean number of participating nucleons \mnpart~as derived from 
the centrality percentile using a Glauber MC calculation~\cite{dndeta}. 
For the results presented as a function of the mean charged-particle density \mdndeta, 
the mean value $\left<N_{\rm acc}\right>$~in each centrality bin is associated with the measured value 
for \mdndeta~from \cite{dndeta}. 
A linear relation between $\langle N_{\rm acc} \rangle$ and \mdndeta~is observed over the full centrality 
range, allowing interpolation to assign a value for \mdndeta~to any interval of $N_{\rm acc}$. 
In pp, \mdndeta~is calculated for each interval of $N_{\rm acc}$ employing the full detector response 
matrix from MC and unfolding of the measured $N_{\rm acc}$ distributions following the 
procedure outlined in~\cite{alice-pp-nch}.

The systematic uncertainties are estimated separately for each collision system (Pb--Pb and pp) 
and at each collision energy.
The relative uncertainties on \rspt~are generally smaller than those on $C_m$ because most of the
sources of uncertainties lead to correlated variations of $M(p_{\rm T})_m$ and $C_m$ that 
tend to cancel in the ratio \rspt. 
Therefore, all quantitative results shown below are presented in terms of \rspt.
The contributions to the total systematic uncertainty on 
\rspt~in pp and Pb--Pb collisions are summarized in Table~\ref{tab:systuncert-rel}. 
Ranges are given when the uncertainties depend on \mdndeta~or centrality.

The largest contribution to the total systematic uncertainty results from the comparison of 
\rspt$~$ from full MC simulations employing a GEANT3~\cite{geant} implementation of the ALICE 
detector setup~\cite{aliroot} to the MC generator level.
Processing the events through the full simulation chain alters the result for \rspt~with respect to the 
MC generator level by up to 6\% in high multiplicity pp collisions. 
This includes effects of tracking efficiency dependence on the transverse momentum. 
The studies in pp are performed using the Perugia-0 tune of PYTHIA6~\cite{skands,pythia}, 
similar results are obtained with PHOJET~\cite{phojet}. 
HIJING~\cite{hijing} is used for Pb--Pb collisions, where the differences are slightly smaller, 
reaching up to 4\% in most central collisions.

Since these deviations are in general dependent on the event characteristics assumed in the model, 
in particular on the nature of the underlying particle correlations, no correction of experimental 
results is performed. Instead, these deviations are added to the systematic data uncertainties to allow 
for a comparison of the experimental results to model calculations on the MC event generator level.

Another major contribution to the total systematic uncertainty emerges from the difference between the 
standard analysis using only TPC tracks and an alternative analysis employing a hybrid tracking scheme.
The hybrid tracking combines TPC and ITS tracks when ITS detector information is available, and thus 
provides more powerful suppression of secondary particles (remaining contamination 4--5\%) as compared
to the standard TPC-only tracking ($\sim$12\%). 
The TPC, on the other hand, 
features very stable operational conditions throughout the analyzed data sets.
The differences between the results from the two analyses reach 5\% in \rspt.

At the event level, minor contributions to the total systematic uncertainty arise from the cut on 
the maximum distance of the reconstructed vertex to the nominal interaction point along the beam axis. 
In the standard analysis global tracks that combine TPC and ITS track segments 
are used for the vertex calculation. 
Alternatively, we studied also the results when only TPC tracks or only
tracklets from the SPD are used to reconstruct the primary vertex. 
The effect from using the different vertex estimators is negligible in Pb--Pb collisions. 
In pp collisions, this effect is small with the exception of the lowest multiplicity bin, where it
reaches 2\% in \rspt.
Additionally, the cut on the difference between the $z$-positions of the reconstructed 
vertices obtained from global tracks and TPC-only tracks is varied. 
This shows a sizable effect only in peripheral Pb--Pb and low-multiplicity pp collisions 
(2--3\% in \rspt).

In addition, variations of the following track quality cuts are performed: 
the number of space points per track in the TPC, 
the $\chi^{2}$ per degree of freedom of the momentum fit, 
and the DCA of each track to the primary vertex, both along the beam direction and in the transverse plane. 
Neither of these contributions to the total systematic uncertainty exceeds 3\% in \rspt.

\begin{table}[h]
	\centering
		\begin{tabular}{|l||c|c|c|c|}
		  \hline
			\textbf{Collision system} & \textbf{pp} & \textbf{pp} & \textbf{pp} & \textbf{Pb--Pb} \\
			\textbf{$\sqrt{s_{\rm NN}}$} & \textbf{0.9~TeV} & \textbf{2.76~TeV} & \textbf{7~TeV} & \textbf{2.76 TeV} \\
			\hline
			\hline
			Vertex $z$-position cut & 0--0.5\% & $<$0.1\% & $<$0.1\% & 0.5--1\% \\
			Vertex calculation & 0--2\% & 0.5--2\% & 0.5--2\% & $<$0.1\% \\
			Vertex difference cut & 0--1.5\% & 0--3\% & 0--2\% & 0--2\% \\
			\hline
			Min.\,TPC space points & 1.5--3\% & 1--2\% & 1--3\% & 2--3\% \\
			TPC $\chi^{2}$ / d.o.f. & $<$0.1\% & $<$0.1\% & $<$0.1\% & $<$0.1\% \\
			DCA to vertex & 1\% & 1--1.5\% & 0.5--1\% & 0.5--1\% \\
			\hline
			B-field polarity & 0.5\% & 0.5\% & 0.5\% & 0.5\% \\
			Centrality intervals & - & - & - & 1--3\% \\
			\hline
			TPC--only vs.\,hybrid~ & 4\% & 4\% & 4\% & 1--5\% \\
			\hline
			MC generator vs.\,full sim. & 0--6\% & 0--6\% & 0--6\% & 0--4\% \\
			\hline
			\hline
			\textbf{Total} & \textbf{4.4--7.7\%} & \textbf{4.4--7.6\%} & \textbf{4.4--7.9\%} & \textbf{4.2--7.4\%} \\
			\hline
		\end{tabular}
	\caption{Contributions to the total systematic uncertainty on $\sqrt{C_{m}} / M(p_{\rm T})_m$ 
	in pp and Pb--Pb collisions. 
	Ranges are given when the uncertainties depend on \mdndeta~or centrality.}
	\label{tab:systuncert-rel}
\end{table}

The difference between the results obtained from Pb--Pb data taken at the two 
magnetic field polarities is included into the systematic uncertainties. 
The effect is small (0.5\% in \rspt).
The corresponding uncertainty in pp is assumed to be the same as in Pb--Pb collisions. 
Finally, the effect of finite centrality intervals in Pb--Pb, 
and the corresponding variation of $M(p_{\rm T})$ within these intervals, 
is taken into account by including the difference between the analyses in 5\% and 10\% centrality 
intervals~\cite{dndeta,centr-new} into the systematic uncertainty. 
The total uncertainty on \rspt~for each data set was obtained by adding in quadrature 
the individual contributions in Table~\ref{tab:systuncert-rel}.

\section{Results in pp collisions}

\begin{figure}[t]
	\centering
		\includegraphics[width=0.49\textwidth]{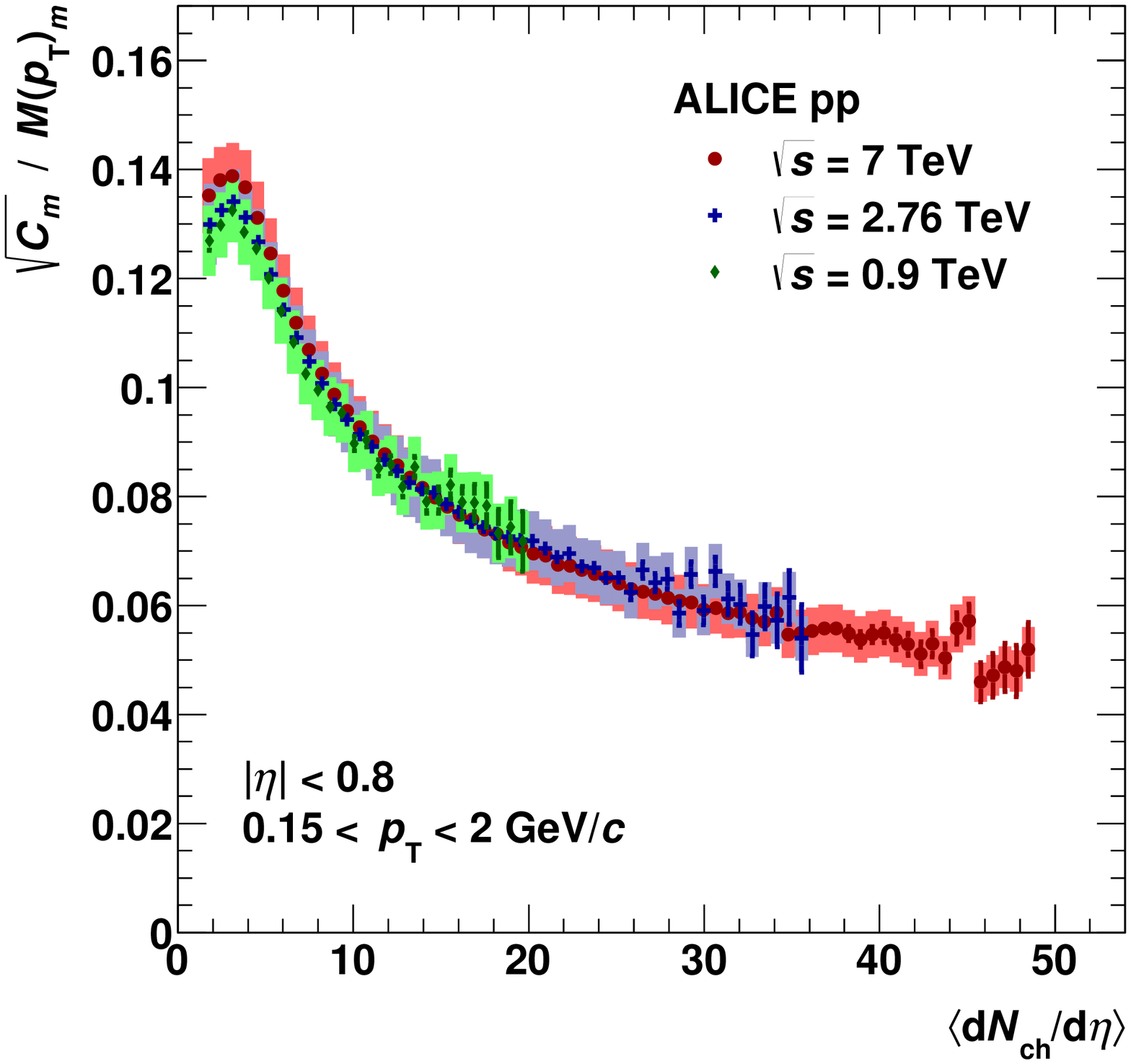}
	\caption{Relative fluctuation \rspt~as a function of \mdndeta~in pp collisions 
	at \roots~$=$~0.9, 2.76 and 7~TeV.}
	\label{fig:pp-corr}
\end{figure}

The relative dynamical fluctuation \rspt~as a function of the average charged-particle 
multiplicity \mdndeta~in pp collisions at \roots~$=$~0.9, 2.76 and 7~TeV 
is shown in Fig.~\ref{fig:pp-corr}. 
The non-zero values of \rspt~indicate significant dynamical 
event-by-event $M(p_{\rm T})$ fluctuations.
The fluctuation strength reaches a maximum of 12--14\% in low multiplicity collisions 
and decreases to about 5\% at the highest multiplicities. 
No significant beam energy dependence is observed for the relative fluctuation \rspt.
 
\begin{figure}
	\centering
		\includegraphics[width=0.49\textwidth]{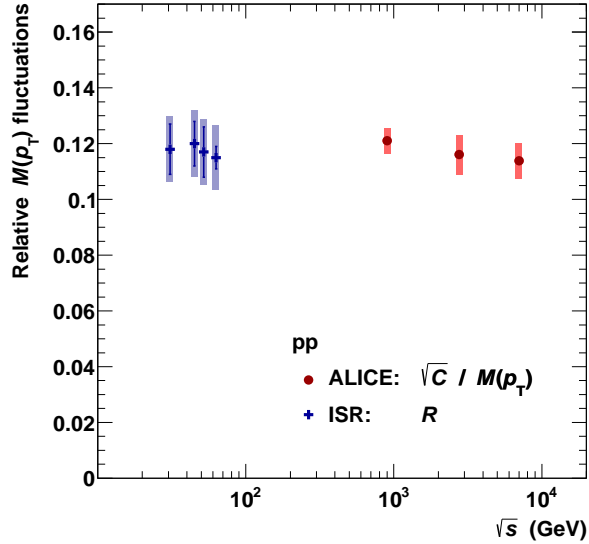}
	\caption{Relative dynamical mean transverse momentum fluctuations
	in pp collisions as a function of \roots. The ALICE results for \rsptinc~are compared
	to the quantity $R$ measured at the ISR (see text and~\cite{isr}).}
	\label{fig:roots}
\end{figure}

The beam energy dependence of relative dynamical mean transverse momentum 
fluctuations in pp collisions was studied at lower collision energies by the 
Split Field Magnet (SFM) detector at the Intersection Storage
Rings (ISR). The SFM experiment measured relative
fluctuations in inclusive pp collisions at \roots~$=$~30.8, 45, 52, and 63 GeV~\cite{isr}.
The fluctuations are expressed by the quantity $R$ that is extracted from the multiplicity dependence 
of the event-by-event $M(p_{\rm T})$~dispersion.
The measure $R=\left[ D(  M_{\rm EbE}(p_{\rm T})_k)/M(p_{\rm T})\right]_{n\rightarrow\infty}$ 
is obtained from an extrapolation of the multiplicity-dependent dispersion $D(  M_{\rm EbE}(p_{\rm T})_k)$ 
to infinite multiplicity, normalized by the inclusive mean transverse momentum. 
It is an alternative approach to extract dynamical transverse momentum fluctuations 
in inclusive pp collisions.

\begin{figure}
	\centering
		\includegraphics[width=0.49\textwidth]{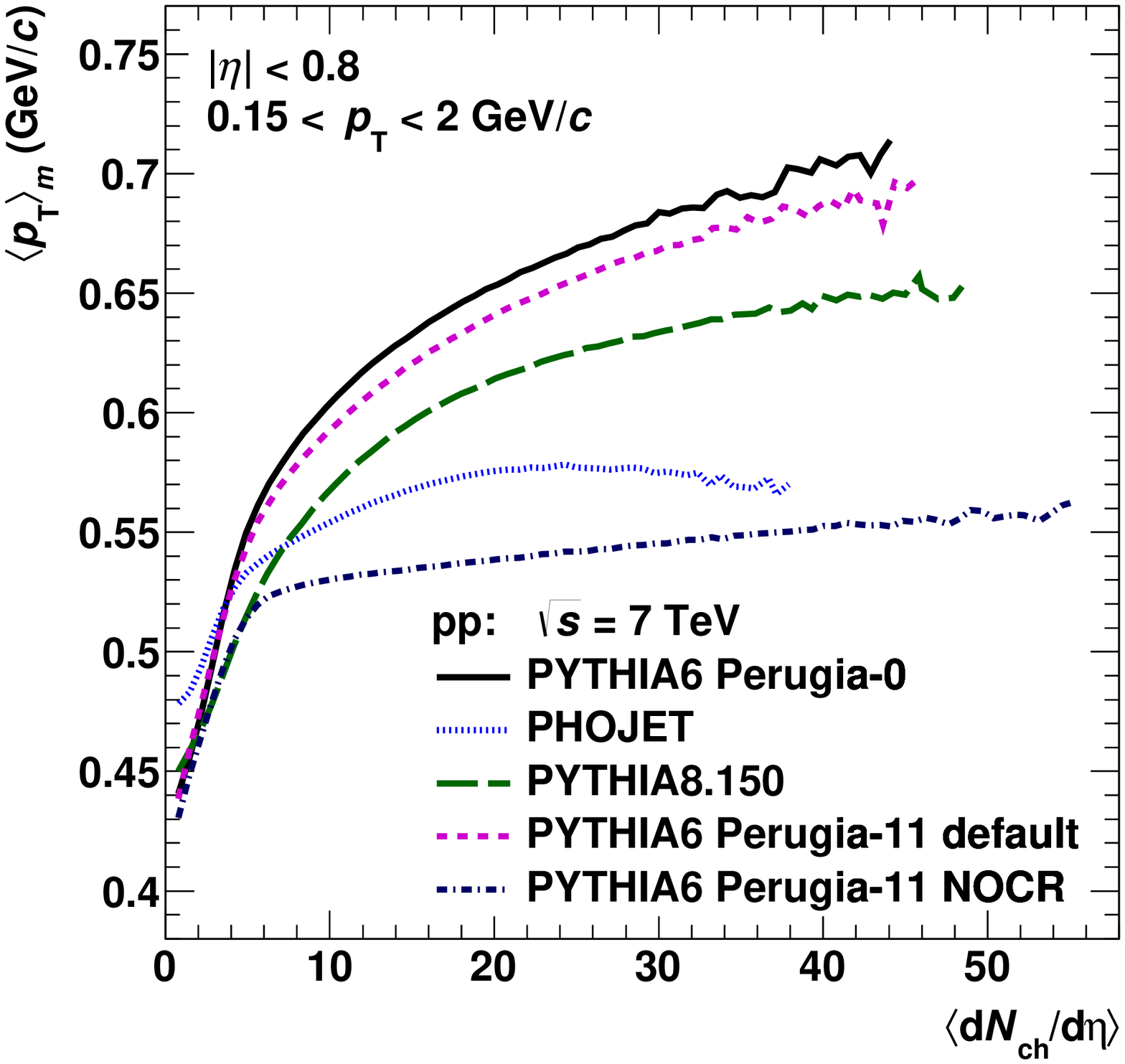}
	\caption{Results for \mpt$_m$ as a function of \mdndeta~in pp collisions at 
	\roots~$=$~7~TeV from different event generators.}
	\label{fig:mpt-pp}
\end{figure}

\begin{figure}
	\centering
		\includegraphics[width=0.49\textwidth]{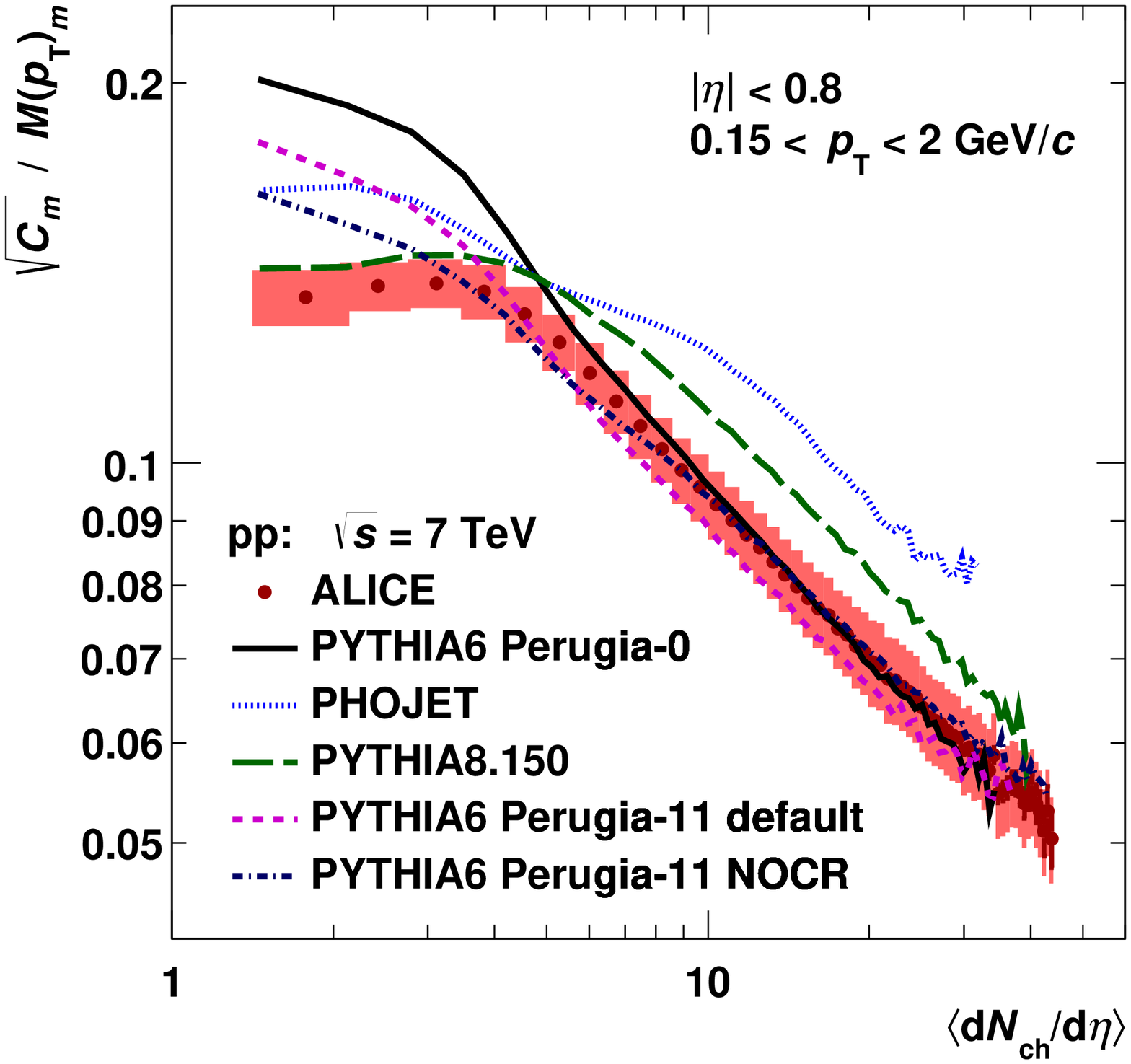}
		\includegraphics[width=0.49\textwidth]{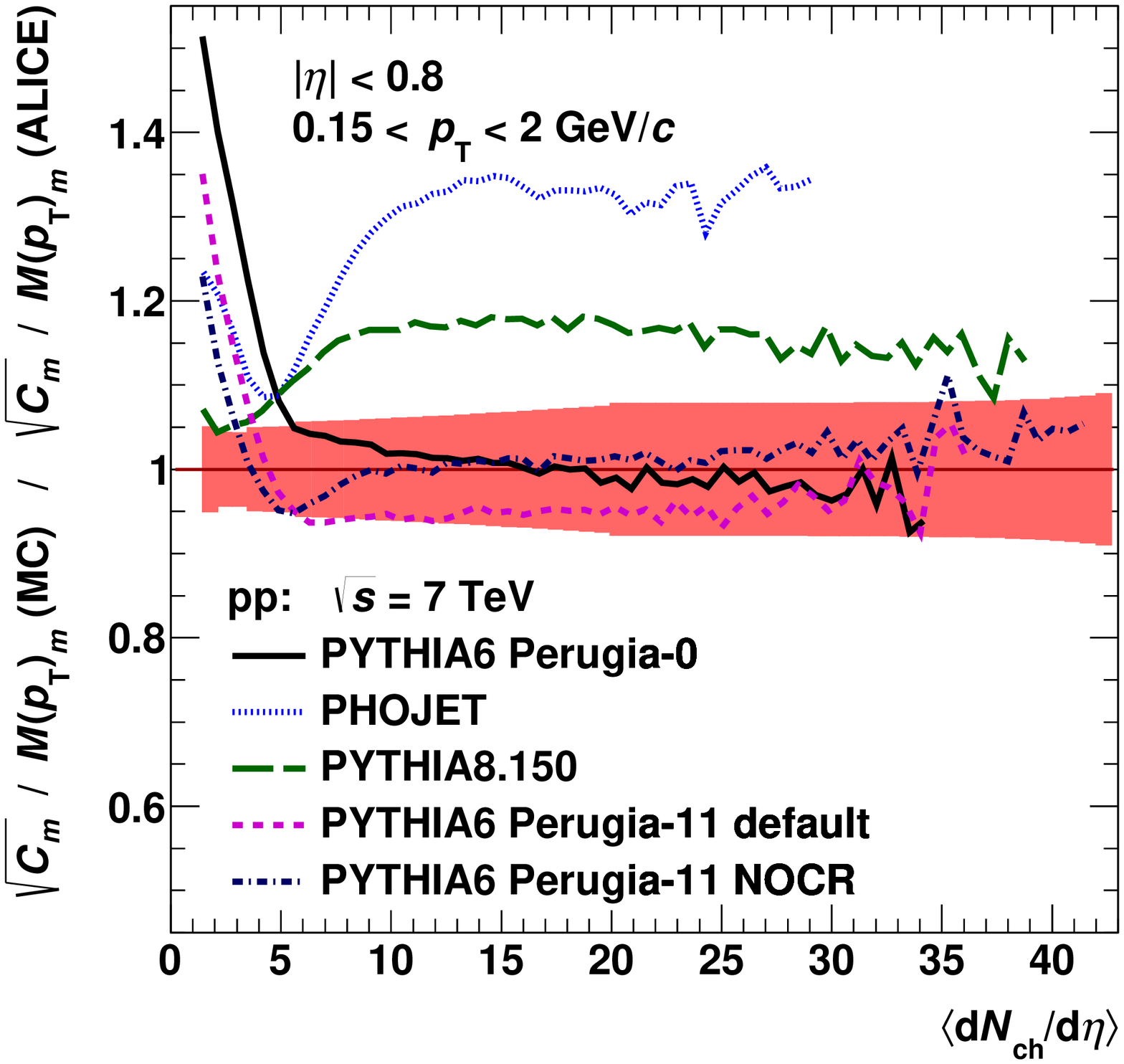}
	\caption{Left: Relative dynamical fluctuation \rspt~for data and different 
	event generators in pp collisions at \roots~$=$~7~TeV as a function of \mdndeta. 
	Right: Ratio models to data. The red error band indicates the statistical and systematic 
	data uncertainties added in quadrature.}
	\label{fig:cm-pp}
\end{figure}

To allow for a comparison to ISR results, an inclusive analysis of ALICE pp data is performed.
The relative fluctuation \rsptinc~is computed at each collision energy as in (\ref{eq:correlator}), 
however, without subdivision into multiplicity classes $m$. 
Monte Carlo studies of pp collisions at \roots~$=$~7~TeV using PYTHIA8 have shown 
that results for R and \rsptinc~agree within 10--15\%. 
The ALICE results for the inclusive \rsptinc~as a function of \roots~are shown 
in Fig.~\ref{fig:roots}, along with the ISR results for $R$ from~\cite{isr}. 
No significant dependence of the relative transverse momentum 
fluctuations on the collision energy is observed over this large energy range. 

The results in pp at \roots~$=$~7~TeV are compared with results from different event generators. 
In particular, PYTHIA6 (tunes Perugia-0 and Perugia-11), PYTHIA8.150 and PHOJET have been used.

It has been pointed out that high-multiplicity events in pp collisions
at LHC energies are driven by multi-parton interactions (MPIs)~\cite{mpi}. 
This picture is also suggested by recent studies of the event sphericity in 
pp collisions~\cite{sphericity}.
MPIs are independent processes on the perturbative level. However, the color reconnection mechanism 
between produced strings may lead to correlations in the hadronic final state. 
Color reconnection is also the driving mechanism in PYTHIA for the increase of \mpt~as a function 
of $N_{\rm ch}$~\cite{cr,mptvsnch}.

The default PYTHIA6 Perugia-11 tune including the color-reconnection mechanism is compared 
to results of the same tune without color-reconnection (NOCR). 
Figure~\ref{fig:mpt-pp} shows model calculations for \mpt$_m$ 
as a function of \mdndeta~in $0.15<p_{\rm T}<2$~GeV/$c$ and $|\eta|<0.8$  
in pp collisions at \roots~$=$~7~TeV. 
The MC generators yield qualitatively different results for the multiplicity dependence, 
in particular PHOJET and the NOCR version of PYTHIA6 Perugia-11 
show only little increase of \mpt$_m$ with multiplicity.
Good agreement between PYTHIA8 and ALICE results in pp collisions at \roots~$=$~7~TeV 
was demonstrated~\cite{mptvsnch}, albeit in a different $\eta$ and $p_{\rm T}$ interval.

Results for the relative dynamical fluctuation measure \rspt~in pp at \roots~$=$~7~TeV are compared 
to model calculations in Fig.~\ref{fig:cm-pp}. The data exhibit a clear power-law dependence 
with \mdndeta~except for very small multiplicities. A power-law fit of \rspt\,$\propto$\,\mdndeta$^{b}$ 
in the interval $5<$\,\mdndeta\,$<30$
yields $b= -0.431 \pm 0.001$\,(stat.)\,$\pm 0.021$\,(syst.). 
The deviation of the power-law index from $b=-0.5$ indicates 
that the observed multiplicity dependence of $M(p_{\rm T})$ fluctuations in pp does not
follow a simple superposition scenario, contrary to what might be expected for independent MPIs.
All PYTHIA tunes under study agree with this finding to the extent that they exhibit a 
similar power-law index as the data. This is also true for the NOCR calculation which 
excludes the color reconnection mechanism in its present implementation in PYTHIA
as a dominant source of correlations beyond the independent superposition scenario.

\section{Results in Pb--Pb collisions}

\begin{figure}[t]
	\centering
		\includegraphics[width=0.49\textwidth]{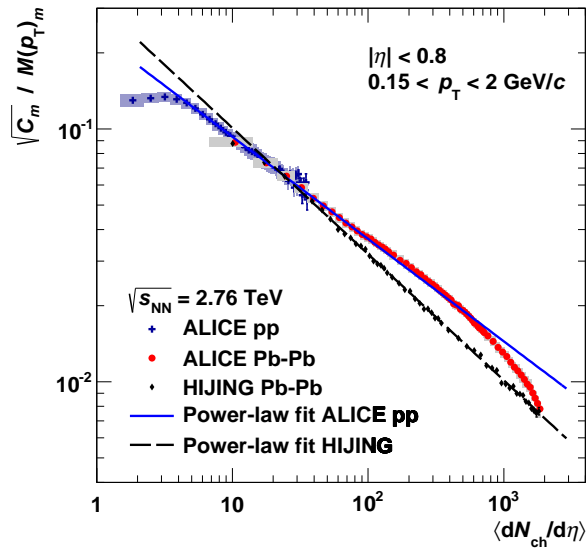}
	\caption{Relative dynamical fluctuation \rspt~as a function of \mdndeta~in pp and Pb--Pb collisions 
	at \rootsnn~$=$~2.76~TeV. Also shown are results from HIJING and power-law fits to pp (solid line) 
	and HIJING (dashed line) (see text).}
	\label{fig:pp-PbPb}
\end{figure}

Results for the relative dynamical fluctuation \rspt~in Pb--Pb collisions at \rootsnn~$=$~2.76~TeV 
as a function of \mdndeta~are shown in Fig.~\ref{fig:pp-PbPb}.
As for pp collisions, significant dynamical fluctuations as well as a strong decrease with 
multiplicity are observed.  
Also shown in Fig.~\ref{fig:pp-PbPb} is the result of a 
HIJING~\cite{hijing} simulation (version 1.36) without jet-quenching.
A power-law fit in the interval $30<$\,\mdndeta\,$<1500$
describes the HIJING results very well, except at low multiplicities, and yields 
$b= -0.499 \pm 0.003$\,(stat.)\,$\pm 0.005$\,(syst.).
The approximate \mdndeta$^{-0.5}$ scaling reflects the basic property of HIJING as a 
superposition model of independent nucleon-nucleon collisions.
The HIJING calculation, in particular the multiplicity dependence, is in obvious
disagreement with the data. 

In peripheral collisions $\left( \left<{\rm d}N_{\rm ch}/{\rm d}\eta\right> < 100 \right)$, 
the Pb--Pb results are in very good agreement 
with the extrapolation of a power-law fit to pp data at \roots~$=$~2.76~TeV 
in the interval $5<$\,\mdndeta\,$<25$, with 
$b= -0.405 \pm 0.002$\,(stat.)\,$\pm 0.036$\,(syst.). 
This is remarkable because significant differences in \mpt~are observed
between pp and Pb--Pb in this multiplicity range~\cite{mptvsnch}.
At larger multiplicities, the Pb--Pb results deviate from the pp extrapolation.
An enhancement in $100<$\,\mdndeta\,$<500$
is followed by a pronounced decrease at \mdndeta\,$>$\,500, 
corresponding to centralities $<$\,40\%, 
which indicates a strong reduction of fluctuations towards central collisions.

\begin{figure}[t]
	\centering
		\includegraphics[width=0.49\textwidth]{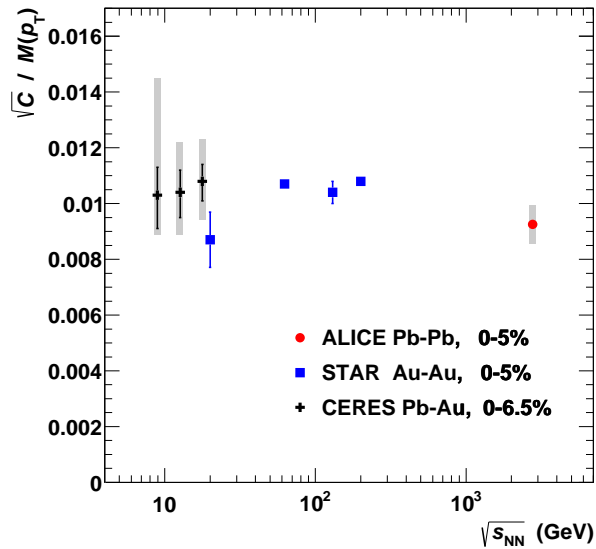}
	\caption{Mean transverse momentum fluctuations in central heavy-ion collisions 
	as a function of \rootsnn. 
	The ALICE data point is compared to data from the CERES~\cite{ceres-2} 
	and STAR~\cite{star-2} experiments. 
	For STAR only statistical uncertainties are available.}
	\label{fig:energy-AA}
\end{figure}

Measurements of mean transverse momentum fluctuations in central A--A collisions 
at the SPS~\cite{ceres-2} and at RHIC~\cite{star-2} are compared to the ALICE result 
in Fig.~\ref{fig:energy-AA}. 
As in pp, there is no significant dependence on \rootsnn~observed over a wide range 
of collision energies.

\begin{figure}[t]
	\centering
		\includegraphics[width=0.9\textwidth]{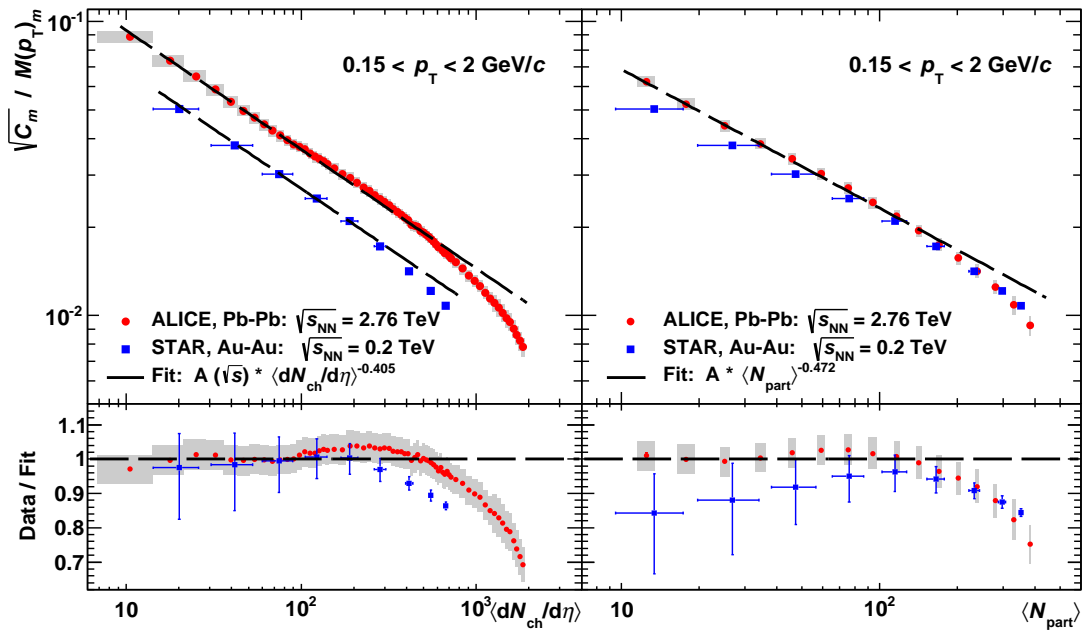}
	\caption{Left: Relative dynamical fluctuation \rspt~as a function of \mdndeta~in Pb--Pb collisions 
	at \rootsnn~$=$~2.76~TeV from ALICE compared to results from STAR in Au--Au collisions 
	at \rootsnn~$=$~200~GeV~\cite{star-2}. 
	Also shown as dashed lines are results from power-law fits to the data (see text). 
	Right: same data as a function of \mnpart.}
	\label{fig:ALICE-STAR}
\end{figure}

Figure~\ref{fig:ALICE-STAR} shows a comparison of the ALICE results for \rspt~to 
measurements in Au--Au collisions at \rootsnn~$=$~200~GeV by the STAR experiment at RHIC~\cite{star-2}.
In the peripheral region, the STAR data show very similar scaling with \mdndeta~as the ALICE data, 
as shown on the left panel of Fig.~\ref{fig:ALICE-STAR}. 
Also shown are the fit to pp data at \roots~$=$~2.76~TeV from Fig.~\ref{fig:pp-PbPb}
and the result of a power-law fit to the STAR data in \mdndeta\,$<$\,200 where the power is fixed
to $b=-0.405$.
Good agreement of the ALICE and STAR data with the fits is observed in peripheral collisions.
The decrease of fluctuations in central collisions is similar in ALICE and STAR, 
however, no significant enhancement in semi-central events is observed in the STAR data.
In the right panel of Fig.~\ref{fig:ALICE-STAR}, the results for \rspt~in ALICE and
STAR are shown as a function of the mean number of participating nucleons \mnpart.
In this representation, the measurements of \rspt~from ALICE and STAR are compatible 
within the rather large experimental uncertainties on \mnpart~in STAR. A power-law fit 
\rspt\,$\propto$\,\mnpart$^{b}$ to
the ALICE data in the interval $10<$\,\mnpart\,$<40$ yields 
$b= -0.472 \pm 0.007$\,(stat.)\,$\pm 0.037$\,(syst.).
The agreement between ALICE and STAR data as a function of \mnpart~points to a relation 
between the observed fluctuation patterns and the collision geometry.

\begin{figure}[t]
	\centering
		\includegraphics[width=0.49\textwidth]{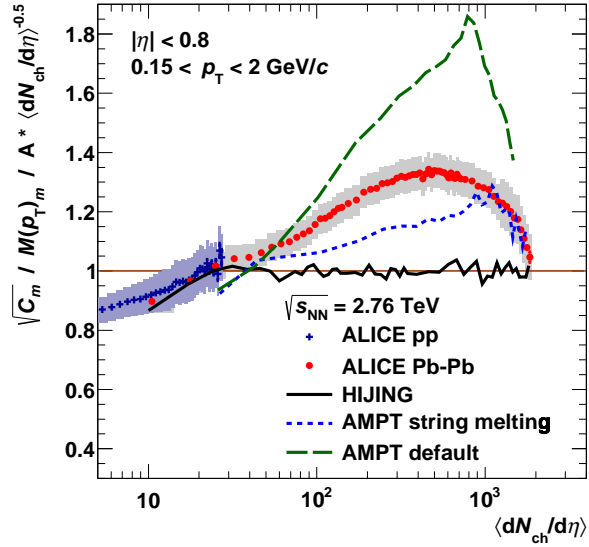}
	\caption{Relative dynamical fluctuation \rspt~normalized to \mdndeta$^{-0.5}$ (see text) 
	as a function of \mdndeta~in pp and Pb--Pb collisions at \rootsnn~$=$~2.76~TeV. 
	The ALICE data are compared to results from HIJING and AMPT.}
	\label{fig:PbPb-models}
\end{figure}

Transverse momentum correlations and fluctuations 
may be modified as a consequence of collective flow in A--A collisions.
It should be noted, however, that event-averaged radial flow and azimuthal asymmetries are
not expected to give rise to strong transverse momentum fluctuations in azimuthally
symmetric detectors~\cite{ceres-2,phenix-2}. On the other hand, $M(p_{\rm T})$ 
fluctuations may occur due to fluctuating initial conditions that are also related 
to event-by-event fluctuations of radial flow and azimuthal asymmetries.
We compare our results to calculations from the AMPT model~\cite{ampt} which has been demonstrated
to give a reasonable description of inclusive and event-averaged bulk properties
in Pb--Pb collisions at LHC energies~\cite{ampt-lhc1,ampt-lhc2}, 
in particular of the measured elliptic flow coefficient $v_2$.
Figure~\ref{fig:PbPb-models} shows the ratio of \rspt~in data and models 
to the result of a fit of $A\cdot$\mdndeta$^{-0.5}$ to the HIJING simulation 
in the interval $30<$\,\mdndeta\,$<1500$.
For \mdndeta\,$<$\,30, HIJING agrees well with the results from pp and Pb--Pb.
At larger multiplicities, none of the models shows quantitative agreement with the Pb--Pb data.
The default AMPT calculation gives rise to increased fluctuations on top of the underlying
HIJING scenario exceeding those observed in the data, except for very peripheral collisions. 
In contrast, the AMPT calculation with string melting,
where partons after rescattering are recombined by a hadronic coalescence scheme, 
predicts smaller fluctuations. On the other hand, both AMPT versions exhibit a pronounced
fall-off in central collisions which is in qualitative agreement with the data.

In a recent approach~\cite{moschelli}, initial spatial fluctuations of glasma flux tubes
have been related to mean transverse momentum fluctuations of final state hadrons
via their coupling to a collective flow field. 
A comparison of these calculations to data from ALICE and STAR is shown in~\cite{moschelli}. 
Good agreement is found in the semi-central and central region, 
where the data deviate from the pp extrapolation.

\section{Summary and conclusions}

First results on event-by-event fluctuations of the mean transverse momentum of charged
particles in pp and Pb--Pb collisions at the LHC are presented. 
Expressed in terms of the relative dynamical fluctuation \rspt, 
little energy dependence of the mean transverse momentum fluctuations is observed in pp at
\roots~$=$~0.9, 2.76 and 7~TeV. The results are also compatible with similar measurements
at the ISR. For the first time, mean transverse momentum fluctuations in pp are
studied as a function of \mdndeta. 
A characteristic decrease of \rspt~following a power law is observed.
The decrease is weaker than expected from a superposition of independent sources. 
The nature of such sources in pp is subject to future studies, but a connection to the 
concept of multi-parton interactions is suggestive.
Model studies using PYTHIA however indicate that there is no strong sensitivity of transverse 
momentum fluctuations to the mechanism of color reconnection.

In peripheral Pb--Pb collisions $\left( \left<{\rm d}N_{\rm ch}/{\rm d}\eta\right> < 100 \right)$, 
the dependence of \rspt~on \mdndeta~is very similar to that observed in pp collisions at the 
corresponding collision energy. 
At larger multiplicities, the Pb--Pb data deviate significantly from an extrapolation
of pp results and show a strong decrease for \mdndeta\,$>$\,500.
The results for the most central collisions are of the same magnitude as previous measurements
at the SPS and at RHIC.
The centrality dependence of \rspt~is compatible with that observed in Au--Au at \rootsnn~$=$~200~GeV.

The Pb--Pb data can not be described by models based on independent nucleon-nucleon collisions
such as HIJING. Models which include initial state density fluctuations and their effect
on the development of collectivity in the final state are in 
qualitative agreement with the data. This suggests a connection between the 
observed fluctuations of transverse momentum and azimuthal
correlations, and their relation to fluctuations in the initial state of the collision.
%
%
\newenvironment{acknowledgement}{\relax}{\relax}
\begin{acknowledgement}
\section*{Acknowledgements}
The ALICE Collaboration would like to thank all its engineers and technicians for their invaluable contributions to the construction of the experiment and the CERN accelerator teams for the outstanding performance of the LHC complex.
\\
The ALICE Collaboration gratefully acknowledges the resources and support provided by all Grid centres and the Worldwide LHC Computing Grid (WLCG) collaboration.
\\
The ALICE Collaboration acknowledges the following funding agencies for their support in building and
running the ALICE detector:
 \\
State Committee of Science,  World Federation of Scientists (WFS)
and Swiss Fonds Kidagan, Armenia,
 \\
Conselho Nacional de Desenvolvimento Cient\'{\i}fico e Tecnol\'{o}gico (CNPq), Financiadora de Estudos e Projetos (FINEP),
Funda\c{c}\~{a}o de Amparo \`{a} Pesquisa do Estado de S\~{a}o Paulo (FAPESP);
 \\
National Natural Science Foundation of China (NSFC), the Chinese Ministry of Education (CMOE)
and the Ministry of Science and Technology of China (MSTC);
 \\
Ministry of Education and Youth of the Czech Republic;
 \\
Danish Natural Science Research Council, the Carlsberg Foundation and the Danish National Research Foundation;
 \\
The European Research Council under the European Community's Seventh Framework Programme;
 \\
Helsinki Institute of Physics and the Academy of Finland;
 \\
French CNRS-IN2P3, the `Region Pays de Loire', `Region Alsace', `Region Auvergne' and CEA, France;
 \\
German BMBF and the Helmholtz Association;
\\
General Secretariat for Research and Technology, Ministry of
Development, Greece;
\\
Hungarian OTKA and National Office for Research and Technology (NKTH);
 \\
Department of Atomic Energy and Department of Science and Technology of the Government of India;
 \\
Istituto Nazionale di Fisica Nucleare (INFN) and Centro Fermi -
Museo Storico della Fisica e Centro Studi e Ricerche "Enrico
Fermi", Italy;
 \\
MEXT Grant-in-Aid for Specially Promoted Research, Ja\-pan;
 \\
Joint Institute for Nuclear Research, Dubna;
 \\
National Research Foundation of Korea (NRF);
 \\
CONACYT, DGAPA, M\'{e}xico, ALFA-EC and the EPLANET Program
(European Particle Physics Latin American Network)
 \\
Stichting voor Fundamenteel Onderzoek der Materie (FOM) and the Nederlandse Organisatie voor Wetenschappelijk Onderzoek (NWO), Netherlands;
 \\
Research Council of Norway (NFR);
 \\
Polish Ministry of Science and Higher Education;
 \\
National Science Centre, Poland;
 \\
 Ministry of National Education/Institute for Atomic Physics and CNCS-UEFISCDI - Romania;
 \\
Ministry of Education and Science of Russian Federation, Russian
Academy of Sciences, Russian Federal Agency of Atomic Energy,
Russian Federal Agency for Science and Innovations and The Russian
Foundation for Basic Research;
 \\
Ministry of Education of Slovakia;
 \\
Department of Science and Technology, South Africa;
 \\
CIEMAT, EELA, Ministerio de Econom\'{i}a y Competitividad (MINECO) of Spain, Xunta de Galicia (Conseller\'{\i}a de Educaci\'{o}n),
CEA\-DEN, Cubaenerg\'{\i}a, Cuba, and IAEA (International Atomic Energy Agency);
 \\
Swedish Research Council (VR) and Knut $\&$ Alice Wallenberg
Foundation (KAW);
 \\
Ukraine Ministry of Education and Science;
 \\
United Kingdom Science and Technology Facilities Council (STFC);
 \\
The United States Department of Energy, the United States National
Science Foundation, the State of Texas, and the State of Ohio;
\\
Ministry of Science, Education and Sports of Croatia and  Unity through Knowledge Fund, Croatia.

\end{acknowledgement}
%
%
\bibliographystyle{utphys}   
\bibliography{pt_fluc_bibliography}
\newpage
%
%
\appendix
\section{The ALICE Collaboration}
\label{app:collab}



\begingroup
\small
\begin{flushleft}
B.~Abelev\Irefn{org71}\And
J.~Adam\Irefn{org37}\And
D.~Adamov\'{a}\Irefn{org79}\And
M.M.~Aggarwal\Irefn{org83}\And
G.~Aglieri~Rinella\Irefn{org34}\And
M.~Agnello\Irefn{org107}\textsuperscript{,}\Irefn{org90}\And
A.~Agostinelli\Irefn{org26}\And
N.~Agrawal\Irefn{org44}\And
Z.~Ahammed\Irefn{org126}\And
N.~Ahmad\Irefn{org18}
% ...
\renewcommand\labelenumi{\textsuperscript{\theenumi}~}

\section*{Affiliation notes}
\renewcommand\theenumi{\roman{enumi}}
\begin{Authlist}
\item \Adef{0}Deceased
\item \Adef{idp1115504}{Also at: St. Petersburg State Polytechnical University}
\item \Adef{idp3044384}{Also at: Department of Applied Physics, Aligarh Muslim University, Aligarh, India}
\item \Adef{idp3724256}{Also at: M.V. Lomonosov Moscow State University, D.V. Skobeltsyn Institute of Nuclear Physics, Moscow, Russia}
\item \Adef{idp3981056}{Also at: University of Belgrade, Faculty of Physics and "Vin\v{c}a" Institute of Nuclear Sciences, Belgrade, Serbia}
\item \Adef{idp4299760}{Permanent Address: Permanent Address: Konkuk University, Seoul, Korea}
\item \Adef{idp4863472}{Also at: Institute of Theoretical Physics, University of Wroclaw, Wroclaw, Poland}
\item \Adef{idp5796336}{Also at: University of Kansas, Lawrence, KS, United States}
\end{Authlist}

\section*{Collaboration Institutes}
\renewcommand\theenumi{\arabic{enumi}~}
\begin{Authlist}

\item \Idef{org1}A.I. Alikhanyan National Science Laboratory (Yerevan Physics Institute) Foundation, Yerevan, Armenia
\item \Idef{org2}Benem\'{e}rita Universidad Aut\'{o}noma de Puebla, Puebla, Mexico
\item \Idef{org3}Bogolyubov Institute for Theoretical Physics, Kiev, Ukraine
\item \Idef{org4}Bose Institute, Department of Physics and Centre for Astroparticle Physics and Space Science (CAPSS), Kolkata, India
\item \Idef{org5}Budker Institute for Nuclear Physics, Novosibirsk, Russia
\item \Idef{org6}California Polytechnic State University, San Luis Obispo, CA, United States
\item \Idef{org7}Central China Normal University, Wuhan, China
\item \Idef{org8}Centre de Calcul de l'IN2P3, Villeurbanne, France
\item \Idef{org9}Centro de Aplicaciones Tecnol\'{o}gicas y Desarrollo Nuclear (CEADEN), Havana, Cuba
\item \Idef{org10}Centro de Investigaciones Energ\'{e}ticas Medioambientales y Tecnol\'{o}gicas (CIEMAT), Madrid, Spain
\item \Idef{org11}Centro de Investigaci\'{o}n y de Estudios Avanzados (CINVESTAV), Mexico City and M\'{e}rida, Mexico
\item \Idef{org12}Centro Fermi - Museo Storico della Fisica e Centro Studi e Ricerche ``Enrico Fermi'', Rome, Italy
\item \Idef{org13}Chicago State University, Chicago, USA
\item \Idef{org14}Commissariat \`{a} l'Energie Atomique, IRFU, Saclay, France
\item \Idef{org15}COMSATS Institute of Information Technology (CIIT), Islamabad, Pakistan
\item \Idef{org16}Departamento de F\'{\i}sica de Part\'{\i}culas and IGFAE, Universidad de Santiago de Compostela, Santiago de Compostela, Spain
\item \Idef{org17}Department of Physics and Technology, University of Bergen, Bergen, Norway
\item \Idef{org18}Department of Physics, Aligarh Muslim University, Aligarh, India
\item \Idef{org19}Department of Physics, Ohio State University, Columbus, OH, United States
\item \Idef{org20}Department of Physics, Sejong University, Seoul, South Korea
\item \Idef{org21}Department of Physics, University of Oslo, Oslo, Norway
\item \Idef{org22}Dipartimento di Fisica dell'Universit\`{a} 'La Sapienza' and Sezione INFN Rome, Italy
\item \Idef{org23}Dipartimento di Fisica dell'Universit\`{a} and Sezione INFN, Cagliari, Italy
\item \Idef{org24}Dipartimento di Fisica dell'Universit\`{a} and Sezione INFN, Trieste, Italy
\item \Idef{org25}Dipartimento di Fisica dell'Universit\`{a} and Sezione INFN, Turin, Italy
\item \Idef{org26}Dipartimento di Fisica e Astronomia dell'Universit\`{a} and Sezione INFN, Bologna, Italy
\item \Idef{org27}Dipartimento di Fisica e Astronomia dell'Universit\`{a} and Sezione INFN, Catania, Italy
\item \Idef{org28}Dipartimento di Fisica e Astronomia dell'Universit\`{a} and Sezione INFN, Padova, Italy
\item \Idef{org29}Dipartimento di Fisica `E.R.~Caianiello' dell'Universit\`{a} and Gruppo Collegato INFN, Salerno, Italy
\item \Idef{org30}Dipartimento di Scienze e Innovazione Tecnologica dell'Universit\`{a} del  Piemonte Orientale and Gruppo Collegato INFN, Alessandria, Italy
\item \Idef{org31}Dipartimento Interateneo di Fisica `M.~Merlin' and Sezione INFN, Bari, Italy
\item \Idef{org32}Division of Experimental High Energy Physics, University of Lund, Lund, Sweden
\item \Idef{org33}Eberhard Karls Universit\"{a}t T\"{u}bingen, T\"{u}bingen, Germany
\item \Idef{org34}European Organization for Nuclear Research (CERN), Geneva, Switzerland
\item \Idef{org35}Faculty of Engineering, Bergen University College, Bergen, Norway
\item \Idef{org36}Faculty of Mathematics, Physics and Informatics, Comenius University, Bratislava, Slovakia
\item \Idef{org37}Faculty of Nuclear Sciences and Physical Engineering, Czech Technical University in Prague, Prague, Czech Republic
\item \Idef{org38}Faculty of Science, P.J.~\v{S}af\'{a}rik University, Ko\v{s}ice, Slovakia
\item \Idef{org39}Frankfurt Institute for Advanced Studies, Johann Wolfgang Goethe-Universit\"{a}t Frankfurt, Frankfurt, Germany
\item \Idef{org40}Gangneung-Wonju National University, Gangneung, South Korea
\item \Idef{org41}Gauhati University, Department of Physics, Guwahati, India
\item \Idef{org42}Helsinki Institute of Physics (HIP), Helsinki, Finland
\item \Idef{org43}Hiroshima University, Hiroshima, Japan
\item \Idef{org44}Indian Institute of Technology Bombay (IIT), Mumbai, India
\item \Idef{org45}Indian Institute of Technology Indore, Indore (IITI), India
\item \Idef{org46}Inha University, Incheon, South Korea
\item \Idef{org47}Institut de Physique Nucl\'eaire d'Orsay (IPNO), Universit\'e Paris-Sud, CNRS-IN2P3, Orsay, France
\item \Idef{org48}Institut f\"{u}r Informatik, Johann Wolfgang Goethe-Universit\"{a}t Frankfurt, Frankfurt, Germany
\item \Idef{org49}Institut f\"{u}r Kernphysik, Johann Wolfgang Goethe-Universit\"{a}t Frankfurt, Frankfurt, Germany
\item \Idef{org50}Institut f\"{u}r Kernphysik, Westf\"{a}lische Wilhelms-Universit\"{a}t M\"{u}nster, M\"{u}nster, Germany
\item \Idef{org51}Institut Pluridisciplinaire Hubert Curien (IPHC), Universit\'{e} de Strasbourg, CNRS-IN2P3, Strasbourg, France
\item \Idef{org52}Institute for Nuclear Research, Academy of Sciences, Moscow, Russia
\item \Idef{org53}Institute for Subatomic Physics of Utrecht University, Utrecht, Netherlands
\item \Idef{org54}Institute for Theoretical and Experimental Physics, Moscow, Russia
\item \Idef{org55}Institute of Experimental Physics, Slovak Academy of Sciences, Ko\v{s}ice, Slovakia
\item \Idef{org56}Institute of Physics, Academy of Sciences of the Czech Republic, Prague, Czech Republic
\item \Idef{org57}Institute of Physics, Bhubaneswar, India
\item \Idef{org58}Institute of Space Science (ISS), Bucharest, Romania
\item \Idef{org59}Instituto de Ciencias Nucleares, Universidad Nacional Aut\'{o}noma de M\'{e}xico, Mexico City, Mexico
\item \Idef{org60}Instituto de F\'{\i}sica, Universidad Nacional Aut\'{o}noma de M\'{e}xico, Mexico City, Mexico
\item \Idef{org61}iThemba LABS, National Research Foundation, Somerset West, South Africa
\item \Idef{org62}Joint Institute for Nuclear Research (JINR), Dubna, Russia
\item \Idef{org63}Konkuk University, Seoul, South Korea
\item \Idef{org64}Korea Institute of Science and Technology Information, Daejeon, South Korea
\item \Idef{org65}KTO Karatay University, Konya, Turkey
\item \Idef{org66}Laboratoire de Physique Corpusculaire (LPC), Clermont Universit\'{e}, Universit\'{e} Blaise Pascal, CNRS--IN2P3, Clermont-Ferrand, France
\item \Idef{org67}Laboratoire de Physique Subatomique et de Cosmologie, Universit\'{e} Grenoble-Alpes, CNRS-IN2P3, Grenoble, France
\item \Idef{org68}Laboratori Nazionali di Frascati, INFN, Frascati, Italy
\item \Idef{org69}Laboratori Nazionali di Legnaro, INFN, Legnaro, Italy
\item \Idef{org70}Lawrence Berkeley National Laboratory, Berkeley, CA, United States
\item \Idef{org71}Lawrence Livermore National Laboratory, Livermore, CA, United States
\item \Idef{org72}Moscow Engineering Physics Institute, Moscow, Russia
\item \Idef{org73}National Centre for Nuclear Studies, Warsaw, Poland
\item \Idef{org74}National Institute for Physics and Nuclear Engineering, Bucharest, Romania
\item \Idef{org75}National Institute of Science Education and Research, Bhubaneswar, India
\item \Idef{org76}Niels Bohr Institute, University of Copenhagen, Copenhagen, Denmark
\item \Idef{org77}Nikhef, National Institute for Subatomic Physics, Amsterdam, Netherlands
\item \Idef{org78}Nuclear Physics Group, STFC Daresbury Laboratory, Daresbury, United Kingdom
\item \Idef{org79}Nuclear Physics Institute, Academy of Sciences of the Czech Republic, \v{R}e\v{z} u Prahy, Czech Republic
\item \Idef{org80}Oak Ridge National Laboratory, Oak Ridge, TN, United States
\item \Idef{org81}Petersburg Nuclear Physics Institute, Gatchina, Russia
\item \Idef{org82}Physics Department, Creighton University, Omaha, NE, United States
\item \Idef{org83}Physics Department, Panjab University, Chandigarh, India
\item \Idef{org84}Physics Department, University of Athens, Athens, Greece
\item \Idef{org85}Physics Department, University of Cape Town, Cape Town, South Africa
\item \Idef{org86}Physics Department, University of Jammu, Jammu, India
\item \Idef{org87}Physics Department, University of Rajasthan, Jaipur, India
\item \Idef{org88}Physik Department, Technische Universit\"{a}t M\"{u}nchen, Munich, Germany
\item \Idef{org89}Physikalisches Institut, Ruprecht-Karls-Universit\"{a}t Heidelberg, Heidelberg, Germany
\item \Idef{org90}Politecnico di Torino, Turin, Italy
\item \Idef{org91}Purdue University, West Lafayette, IN, United States
\item \Idef{org92}Pusan National University, Pusan, South Korea
\item \Idef{org93}Research Division and ExtreMe Matter Institute EMMI, GSI Helmholtzzentrum f\"ur Schwerionenforschung, Darmstadt, Germany
\item \Idef{org94}Rudjer Bo\v{s}kovi\'{c} Institute, Zagreb, Croatia
\item \Idef{org95}Russian Federal Nuclear Center (VNIIEF), Sarov, Russia
\item \Idef{org96}Russian Research Centre Kurchatov Institute, Moscow, Russia
\item \Idef{org97}Saha Institute of Nuclear Physics, Kolkata, India
\item \Idef{org98}School of Physics and Astronomy, University of Birmingham, Birmingham, United Kingdom
\item \Idef{org99}Secci\'{o}n F\'{\i}sica, Departamento de Ciencias, Pontificia Universidad Cat\'{o}lica del Per\'{u}, Lima, Peru
\item \Idef{org100}Sezione INFN, Bari, Italy
\item \Idef{org101}Sezione INFN, Bologna, Italy
\item \Idef{org102}Sezione INFN, Cagliari, Italy
\item \Idef{org103}Sezione INFN, Catania, Italy
\item \Idef{org104}Sezione INFN, Padova, Italy
\item \Idef{org105}Sezione INFN, Rome, Italy
\item \Idef{org106}Sezione INFN, Trieste, Italy
\item \Idef{org107}Sezione INFN, Turin, Italy
\item \Idef{org108}SSC IHEP of NRC Kurchatov institute, Protvino, Russia
\item \Idef{org109}SUBATECH, Ecole des Mines de Nantes, Universit\'{e} de Nantes, CNRS-IN2P3, Nantes, France
\item \Idef{org110}Suranaree University of Technology, Nakhon Ratchasima, Thailand
\item \Idef{org111}Technical University of Split FESB, Split, Croatia
\item \Idef{org112}The Henryk Niewodniczanski Institute of Nuclear Physics, Polish Academy of Sciences, Cracow, Poland
\item \Idef{org113}The University of Texas at Austin, Physics Department, Austin, TX, USA
\item \Idef{org114}Universidad Aut\'{o}noma de Sinaloa, Culiac\'{a}n, Mexico
\item \Idef{org115}Universidade de S\~{a}o Paulo (USP), S\~{a}o Paulo, Brazil
\item \Idef{org116}Universidade Estadual de Campinas (UNICAMP), Campinas, Brazil
\item \Idef{org117}University of Houston, Houston, TX, United States
\item \Idef{org118}University of Jyv\"{a}skyl\"{a}, Jyv\"{a}skyl\"{a}, Finland
\item \Idef{org119}University of Liverpool, Liverpool, United Kingdom
\item \Idef{org120}University of Tennessee, Knoxville, TN, United States
\item \Idef{org121}University of Tokyo, Tokyo, Japan
\item \Idef{org122}University of Tsukuba, Tsukuba, Japan
\item \Idef{org123}University of Zagreb, Zagreb, Croatia
\item \Idef{org124}Universit\'{e} de Lyon, Universit\'{e} Lyon 1, CNRS/IN2P3, IPN-Lyon, Villeurbanne, France
\item \Idef{org125}V.~Fock Institute for Physics, St. Petersburg State University, St. Petersburg, Russia
\item \Idef{org126}Variable Energy Cyclotron Centre, Kolkata, India
\item \Idef{org127}Vestfold University College, Tonsberg, Norway
\item \Idef{org128}Warsaw University of Technology, Warsaw, Poland
\item \Idef{org129}Wayne State University, Detroit, MI, United States
\item \Idef{org130}Wigner Research Centre for Physics, Hungarian Academy of Sciences, Budapest, Hungary
\item \Idef{org131}Yale University, New Haven, CT, United States
\item \Idef{org132}Yonsei University, Seoul, South Korea
\item \Idef{org133}Zentrum f\"{u}r Technologietransfer und Telekommunikation (ZTT), Fachhochschule Worms, Worms, Germany
\end{Authlist}
\endgroup

%
%
\end{document}